\documentclass[preprint, superscriptaddress, preprintnumbers,amsmath,amssymb]{revtex4}

\allowdisplaybreaks
\allowdisplaybreaks[4]
\usepackage{graphicx}% Include figure files
\usepackage{dcolumn}% Align table columns on decimal point
\usepackage{bm}% bold math
\usepackage[dvipdfm,
            pdfstartview=FitH,
            CJKbookmarks=true,
            bookmarksnumbered=true,
            bookmarksopen=true,
            colorlinks,
            pdfborder=001,
            linkcolor=blue,
            anchorcolor=blue,
            citecolor=blue
            ]{hyperref}
\begin{document}

%\begin{CJK}{GBK}{song}

\title{Chiral geometry in symmetry restored states: Chiral doublet bands in $^{128}$Cs}

\author{F. Q. Chen  }
\affiliation{State Key Laboratory of Nuclear Physics and Technology, School of Physics, Peking University,
Beijing 100871, People's Republic of China}
\author{Q. B. Chen  }
\affiliation{State Key Laboratory of Nuclear Physics and Technology, School of Physics, Peking University,
Beijing 100871, People's Republic of China}
\author{Y. A. Luo  }
\affiliation{School of Physics, Nankai University, Tianjin 300071, People's Republic of China}
\author{J. Meng }
 \email{mengj@pku.edu.cn}
\affiliation{State Key Laboratory of Nuclear Physics and Technology, School of Physics, Peking University,
Beijing 100871, People's Republic of China}
\affiliation{School of Physics and Nuclear Energy Engineering, Beihang University, Beijing 100191, People's Republic of China}
%\affiliation{Department of Physics, University of Stellenbosch, Stellenbosch, South Africa}
\author{S. Q. Zhang  }
 \email{sqzhang@pku.edu.cn}
\affiliation{State Key Laboratory of Nuclear Physics and Technology, School of Physics, Peking University,
Beijing 100871, People's Republic of China}

\date{\today}

\begin{abstract}
 The pairing-plus-quadrupole Hamiltonian is diagonalized in a symmetry restored basis, i.e., the triaxial quasiparticle-states with angular momentum and particle number projections, and applied for chiral doublet bands in $^{128}$Cs. The observed energy spectra and electromagnetic transition probabilities are reproduced well without introducing any parameter. The orientation of the angular momentum in the intrinsic frame is investigated by the distributions of its components on the three principle axes as well as those of its tilted angles. The evolution of the chirality with spin is illustrated and the chiral geometry is demonstrated in the angular momentum projected model for the first time.\\
 \\
 keywords: Chirality, Angular momentum projection, Angular momentum geometry, $^{128}$Cs

\end{abstract}

\maketitle

Spontaneous symmetry breaking, in particular spontaneous chiral symmetry breaking, is a subject of general
interest. Spontaneous chiral symmetry breaking in atomic nuclei has attracted intensive investigation since its first prediction by Frauendorf and Meng in 1997 \cite{TAC1997}.
Due to the existence of high-$j$ proton (neutron) particle(s) and neutron (proton) hole(s), the nuclear triaxial shape, and the couplings between single-particle and collective motions, the total nuclear angular momentum vector may lie outside the
three principal planes in the intrinsic frame, leading to the spontaneous chiral symmetry breaking. The experimental signal of the spontaneous chiral symmetry breaking is the observation of the chiral doublet bands, which are a pair of near degenerate $\Delta I=1$ bands with the same parity.
The evidence of the chiral doublet bands was first observed in 2001 \cite{Starosta2001}.
So far,
around 40 candidates of chiral doublet bands have been reported in $A\sim80$ \cite{Br80exp,Br78exp}, $100$ \cite{Rh104exp,Ag106exp1,Rh102exp,Ag106exp3,Ag106exp2,Rh103exp}, $130$ \cite{Starosta2001,Nd135exp,Pr134exp1,Pr134exp2,Cs128-2006,Nd135exp2,Ce133exp} and $190$ \cite{Ir188exp,Tl198exp} mass regions experimentally, see \cite{Meng2010JPG,Meng2014IJMPE,Meng2016PS} for reviews.
Theoretically, the
chiral doublet bands are studied by the particle rotor model
(PRM) \cite{TAC1997,Pengjing2003,Koike2004,ZSQ2007,QB2009PLB}, the titled axis
cranking (TAC) model \cite{TAC1997,Dimitrov2000TAC,Madokoro2000,SkyrmeTAC1,SkyrmeTAC2}, the TAC plus random phase approximation \cite{RPA2011}, the collective Hamiltonian \cite{CQB2013,CQB2016},
and the interacting boson-fermion-fermion model (IBFFM) \cite{Tonev2006,Tonev2007,Brant2008}.

The angular momentum projection (AMP) approach restores the rotational symmetry spontaneously broken in the mean field approximation, and combines the advantages of the TAC model and the PRM. One of the AMP approaches is the projected shell model \cite{Hara1995,Sun1996375,Hara1999,Sunyang2016}, and its applications to triaxial nuclei coupled with quasiparticles can be found in Refs. \cite{GZC2006,Sheikh2008,Bhat2012,Bhat2014}. The attempts to understand the chiral doublet bands by the projected shell model have been made in Refs. \cite{Bhat2012,Bhat2014}. However, it is a big challenge to examine the chiral geometry of angular momentum in the AMP approach due to the complication that projected
basis is defined in the laboratory frame and forms a
non-orthogonal set.

In this Letter, the chiral geometry of angular momentum is investigated within the framework of the AMP approach. The geometry of the angular momentum is analysed in terms of the distributions of its components on the three intrinsic axes (the $K$-distributions), as well as the distribution of its tilted angles in the intrinsic frame, which are calculated within the framework of the AMP for the first time. The typical chiral nucleus $^{128}$Cs \cite{Cs128-2006} is investigated as an example.

As a starting point, we adopt the standard pairing-plus-quadrupole Hamiltonian \cite{ManyBody},
\begin{equation}\label{Hamiltonian}
\hat{H}=\hat{H}_0-\frac{\chi}{2}\sum^2_{\mu=-2}\hat{Q}^+_\mu\hat{Q}_\mu-G_M\hat{P}^+\hat{P}-G_Q\sum^2_{\mu=-2}\hat{P}^+_\mu\hat{P}_\mu,
\end{equation}
which includes the spherical single-particle Hamiltonian, the quadrupole-quadrupole interaction, as well as the monopole and quadrupole pairing. The intrinsic state $|\Phi\rangle$ is obtained by the constrained calculation,
\begin{equation}\label{variational}
\delta\langle\Phi|\hat{H}-\lambda_n\hat{N}-\lambda_p\hat{Z}-\lambda_{q_0}\hat{Q}_{0}-\lambda_{q_2}\hat{Q}_{2}|\Phi\rangle=0,
\end{equation}
with $\lambda_n$ and $\lambda_p$ determined by the particle numbers $N$ and $Z$, and $\lambda_{q_0}$ and $\lambda_{q_2}$ by the quadrupole moments. For odd-odd nuclei, we use $|\Phi_{\nu\pi}\rangle$ to denote the intrinsic state with the neutron (proton) single-particle orbital $\nu$ ($\pi$) blocked. The space of the intrinsic states is spanned by the states $|\Phi_\kappa\rangle\in\{|\Phi_{\nu\pi}\rangle,|\Phi_{\nu\bar{\pi}}\rangle,|\Phi_{\bar{\nu}\pi}\rangle,|\Phi_{\bar{\nu}\bar{\pi}}\rangle\}$,
in which $\bar{\nu}$, $\bar{\pi}$ represent the time-reversal conjugates of $\nu$, $\pi$.

The Hamiltonian (\ref{Hamiltonian}) is diagonalized in the symmetries restored basis obtained by projection approach,
\begin{equation}\label{projectedbasis}
\{\hat{P}^I_{MK}|\Phi_{N,Z,\kappa}\rangle\}\equiv\{\hat{P}^I_{MK}\hat{P}^N\hat{P}^Z|\Phi_{\kappa}\rangle\},
\end{equation}
with the three-dimensional angular momentum projector $\hat{P}^I_{MK}$ and
the particle number projectors $\hat{P}^N$ and $\hat{P}^Z$ \cite{ManyBody}. The corresponding eigen functions are,
\begin{equation}\label{wavefunction}
\begin{split}
|\Psi_{IM}\rangle=&\sum_{K,\kappa} f^I_{K,\kappa}\hat{P}^I_{MK}|\Phi_{N,Z,\kappa}\rangle\\
=&\sum_{K,\kappa}f^I_{K,\kappa}\frac{2I+1}{8\pi^2}\int d\Omega D^{I*}_{MK}(\Omega)\hat{R}(\Omega)|\Phi_{N,Z,\kappa}\rangle,
\end{split}
\end{equation}
in which the coefficient $f^I_{K,\kappa}$ is determined by the Hill-Wheeler equation
\begin{equation}
\sum_{K'\kappa'}(\langle\Phi_{N,Z,\kappa}|\hat{H}\hat{P}^I_{KK'}|\Phi_{N,Z,\kappa'}\rangle
-E_{I}\langle\Phi_{N,Z,\kappa}|\hat{P}^I_{KK'}|\Phi_{N,Z,\kappa'}\rangle)f^{I}_{K'\kappa'}=0.
\end{equation}

It is noted that the information of the orientation of the angular momentum in the intrinsic frame is carried by the quantity $K$. If one regards $\{K,\kappa\}$ in Eq. (\ref{wavefunction}) as the generator coordinates, the corresponding collective wave function $g^I(K,\kappa)$ writes \cite{ManyBody, ErGCM2017},
\begin{equation}\label{collectiveK}
g^I(K,\kappa)=\sum_{K',\kappa'}\mathcal{N}^{1/2}_I(K,\kappa;K',\kappa')f^I_{K',\kappa'},
\end{equation}
with the norm matrix $\mathcal{N}_{I}(K,\kappa;K',\kappa')\equiv\langle\Phi_{N,Z,\kappa}|\hat{P}^I_{KK'}|\Phi_{N,Z,\kappa'}\rangle$. The $K$-distribution for the angular momentum is,
\begin{equation}\label{pK}
p^I(|K|)=\sum_{\kappa}|g^I(K,\kappa)|^2+|g^I(-K,\kappa)|^2,
\end{equation}
which gives the distribution of the long ($l$-) axis component of the angular momentum for triaxiality parameter $\gamma\in[0^\circ, 60^\circ]$. The distributions with respect to the short ($s$-) and the intermediate ($i$-) axes are obtained by replacing $\gamma\in[0^\circ, 60^\circ]$ with $\gamma\in[120^\circ, 180^\circ]$ and $[240^\circ, 300^\circ]$, respectively \cite{QB2009PLB}.

\begin{figure}[!h]
  \begin{center}
    \includegraphics[width=9 cm]{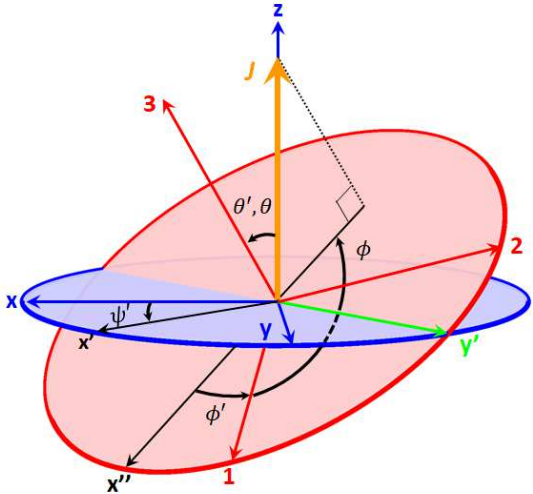}
    \caption{(Color online) Illustration of the relation between the tilted angles ($\theta$, $\phi$) of the angular momentum with respect to the intrinsic frame (labeled by $1$, $2$, $3$) and the Euler angles $\{\psi',\theta',\phi'\}$ describing the orientation of the intrinsic frame with respect to the laboratory frame (labeled by $x$, $y$, $z$).
    }\label{figangles}
  \end{center}
\end{figure}

The angular momentum geometry can also be illustrated by its profile on the ($\theta$, $\phi$) plane, where ($\theta$, $\phi$) are the tilted angles of the angular momentum with respect to the intrinsic frame. For $\gamma\in[0^\circ,60^\circ]$, $\theta$ is the angle between the angular momentum and the $l$-axis, while $\phi$ is the angle between the projection of the angular momentum on the $i$-$s$ plane and the $i$-axis. We found that the profiles can be obtained from the relation between the tilted angles ($\theta$, $\phi$) and the Euler angles $\Omega\equiv\{\psi',\theta',\phi'\}$,
\begin{equation}\label{angles}
\theta=\theta',~~~~\phi=\pi-\phi',
\end{equation}
where the $z$-axis in the laboratory frame is chosen along the angular momentum, as illustrated in Fig. \ref{figangles}.

For the eigen functions (\ref{wavefunction}), if we choose $\{\Omega,\kappa\}$ as the generator coordinates, the corresponding generating function is $\hat{R}(\Omega)|\Phi_{N,Z,\kappa}\rangle$, and the weight function is $F(\Omega,\kappa)\equiv f^I_{K,\kappa}\frac{2I+1}{8\pi^2}D^{I*}_{MK}(\Omega)$. The collective wave function $G^I(\Omega,\kappa)$ is,
\begin{equation}
G^I(\Omega,\kappa)=\sum_{\kappa'}\int d\Omega' \mathcal{M}^{1/2}(\Omega,\kappa;\Omega',\kappa')F(\Omega',\kappa')\\
=\sqrt{\frac{2I+1}{8\pi^2}}\sum_{K}g^I(K,\kappa)D^{I*}_{MK}(\Omega),
\end{equation}
in which  $\mathcal{M}(\Omega,\kappa;\Omega',\kappa')\equiv
\langle\Phi_{N,Z,\kappa}|\hat{R}^+(\Omega)\hat{R}(\Omega')|\Phi_{N,Z,\kappa'}\rangle$, and the relation
\begin{equation}\label{normOmega}
\mathcal{M}^{1/2}(\Omega,\kappa;\Omega',\kappa')=
\sum_{IM}\sum_{KK'}\sqrt{\frac{2I+1}{8\pi^2}}\mathcal{N}^{1/2}_{I}(K,\kappa;K',\kappa')
D^{I*}_{MK}(\Omega)D^{I}_{MK'}(\Omega')
\end{equation}
has been used.

Therefore, the profile for the orientation of the angular momentum in the ($\theta$, $\phi$) plane write,
\begin{equation}
\mathcal{P}(\theta,\phi)=\sum_{\kappa}\int^{2\pi}_{0}d\psi'|G(\psi',\theta,\pi-\phi,\kappa)|^2.
\end{equation}

In the following, the doublet bands in $^{128}$Cs \cite{Cs128-2006} are investigated as an example by the AMP approach. The parameters in the Hamiltonian (\ref{Hamiltonian}) are taken from Ref. \cite{GZC2006}. In the calculation, the quadrupole deformation parameters ($\beta$, $\gamma$) are constrained to be ($0.20$, $30.0^\circ$). This choice of ($\beta$, $\gamma$) is close to the ground state deformation ($0.23$, $23.8^\circ$) \cite{Cs128RMF} given by the covariant density functional theory (CDFT) \cite{Ring1996PPNP,Vretenar2005PR,Meng2006PPNP,CDFT2016} with interaction PC-PK1 \cite{PC-PK1}. It agrees reasonably with the deformations ($0.20$, $37.0^\circ$) used in the projected shell model calculation \cite{GZC2006}. The yrast band is obtained by blocking the lowest $\pi h_{11/2}$ and the fourth $\nu h_{11/2}$ orbitals, which is consistent with the CDFT results \cite{Cs128RMF}.

\begin{figure}[!h]
  \begin{center}
    \includegraphics[width=9 cm]{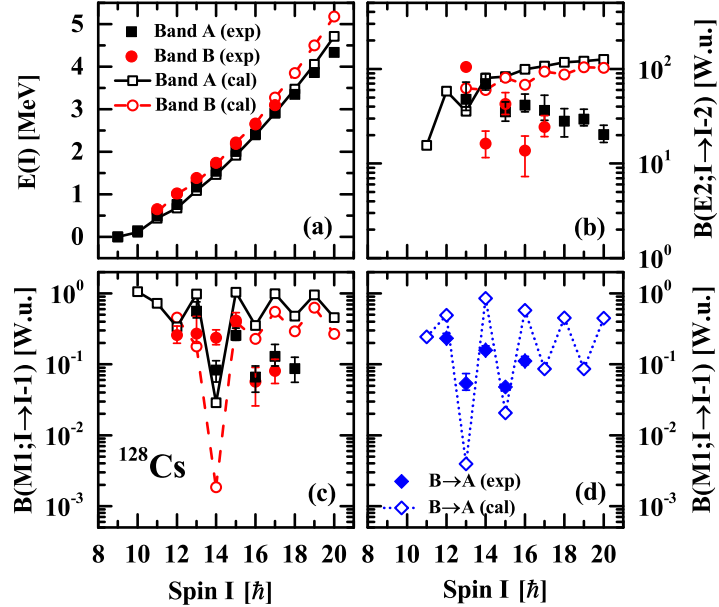}
    \caption{(Color online) The energy spectra (a), the intra-band $B(E2)$ (b) and $B(M1)$ (c), and the inter-band $B(M1)$ (d) of the doublet bands in $^{128}$Cs calculated by the AMP approach in comparison with the data available \cite{Cs128-2006}.}\label{observables}
  \end{center}
\end{figure}

The calculated energy spectra, the intra-band $B(E2)$ and $B(M1)$, and the inter-band $B(M1)$ of the doublet bands (denoted as band A and band B, respectively) in $^{128}$Cs are shown in Fig. \ref{observables}, in comparison with the available data \cite{Cs128-2006}.

The observed energy spectra are excellently reproduced, as shown in Fig. \ref{observables}(a), including the energy difference between the partner bands.

The similarity of the $B(E2)$ between bands A and B, which qualified $^{128}$Cs as the best example of chiral nucleus, is found in both the data and the calculated results, as shown in Fig. \ref{observables}(b). The strength of the calculated $B(E2)$ agrees with the data near the band head. With the increasing spin, the trend of the calculated results deviates from the data due to the frozen nuclear shape. The quality of agreement is comparable with that of the core-quasiparticle coupling model \cite{Koike2003,Cs128-2006}.

The staggering of the intra- and the inter-band $B(M1)$, another signature of chiral modes \cite{Koike2004,WSY2007}, can be seen in both the data and the calculated results, as shown in Figs. \ref{observables}(c) and (d).

As the main features of chiral bands in $^{128}$Cs \cite{Cs128-2006} are well reproduced by the present calculation, it is interesting to study the evolution of the angular momentum geometry and its chiral geometry.

\begin{figure}[!h]
  \begin{center}
    \includegraphics[width=9 cm]{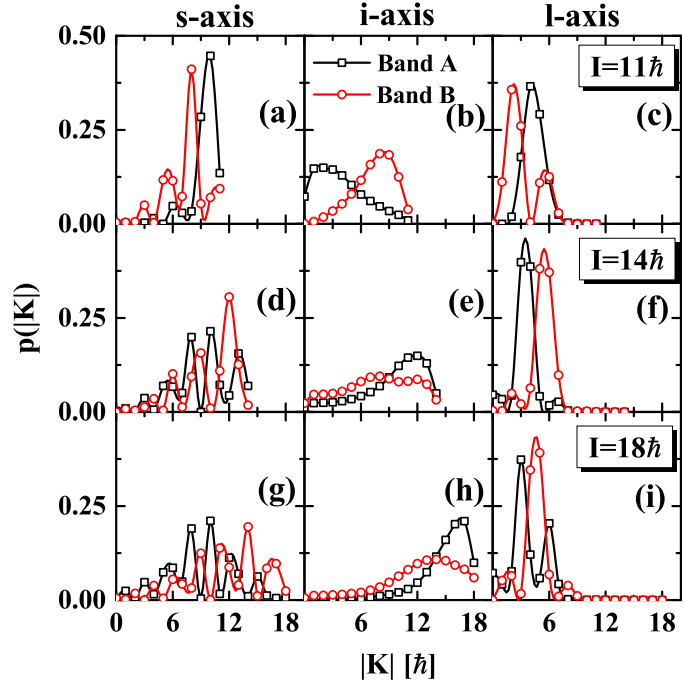}
    \caption{(Color online) $K$-distributions for the angular momentum on the short ($s$), intermediate ($i$) and long ($l$) axes, calculated at $I=11$, $14$ and $18\hbar$, respectively.}\label{Kdistribution}
  \end{center}
\end{figure}

\begin{figure}[!h]
  \begin{center}
    \includegraphics[width=10cm]{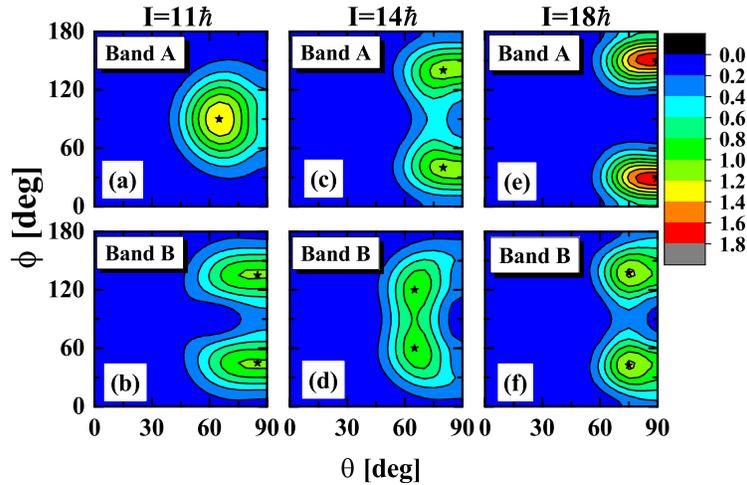}
    \caption{(Color online) Profile for the orientation of the angular momentum in the ($\theta$, $\phi$) plane calculated at $I=11$, $14$ and $18\hbar$, respectively.}\label{tiltedangles}

  \end{center}
\end{figure}

In Fig. \ref{Kdistribution}, the calculated $K$-distribution $p^I(|K|)$ (\ref{pK}) for the doublet bands in $^{128}$Cs
are shown for $I=11$, $14$ and $18\hbar$. Similar discussion has been done extensively within the framework of PRM \cite{QB2009PLB,QB2009PRC}. As seen in the figure, the evolution of chiral modes from the chiral vibration
near the band head to the static chirality at higher spins is exhibited.

For $I=11\hbar$, the probability at $K_i=0$ is significant for band A, while it vanishes for band B. This is in accordance with the interpretation of chiral vibration with respect to the $s$-$l$ plane, where the 0-phonon state (Band A) is symmetric with respect to $K_i=0$ and the 1-phonon state (Band B) is antisymmetric \cite{QB2009PLB,QB2009PRC}.

For $I=14\hbar$, the maximum probability for $K_i$ appears at $K_i\sim12\hbar$ for band A, which means, with the increase of spin, the orientation of the angular momentum deviates from the $s$-$l$ plane and aplanar rotation occurs. The $K$-distributions of band B are similar to those of band A, which indicates the appearance of static chirality. The finite values of $p(K_i = 0)$ and $p(K_l = 0)$ reflect
the tunneling between the left- and right-handed configurations, which is responsible for the energy difference remaining between bands A and B.

For $I=18\hbar$, the
energy difference between the chiral partners increases and the similarity between the $K$-distributions of the two bands becomes less pronounced, which weakens the feature of static chirality.

In Fig. \ref{tiltedangles}, the profiles $\mathcal{P}(\theta,\phi)$ of the doublet bands in $^{128}$Cs are shown for $I=11, 14$ and $18\hbar$ in order to examine the orientation of the angular momentum in the ($\theta$, $\phi$) plane.

For $I=11\hbar$, the angular momentum for band A mainly orientates at ($\theta\sim60^\circ$, $\phi=90^\circ$), namely a planar rotation within the $s$-$l$ plane. The angular momentum for band B orientates equally at ($\theta\sim85^\circ$, $\phi\sim45^\circ$) and ($\theta\sim85^\circ$, $\phi\sim135^\circ$), in accordance with the interpretation of chiral vibration along the $\phi$ direction (i.e., with respect to the $s$-$l$ plane). Here, $\theta\sim85^\circ$ means that the angular momentum is located very close to the $i$-$s$ plane, which is consistent  with the $K_l$-distribution shown in Fig. \ref{Kdistribution} (c), the peak at $K_l=2\hbar$.

For $I=14\hbar$, the angular momenta orientate equally at two directions, for bands A (($\theta\sim80^\circ,\phi\sim40^\circ$) and ($\theta\sim80^\circ,\phi\sim140^\circ$)) and B (($\theta\sim65^\circ,\phi\sim60^\circ$) and ($\theta\sim65^\circ,\phi\sim120^\circ$)), which demonstrates the occurrence of static chirality. The tunneling between the left- and right-handed configurations is reflected by the non-vanishing probability for either $\phi=90^\circ$ or $\theta=90^\circ$, as well as the fact that the orientations of the angular momentum of band A do not coincide exactly with that of band B.

For $I=18\hbar$, the static chirality disappears. The angular momenta for band A orientate to ($\theta=90^\circ, \phi=30^\circ$) and ($\theta=90^\circ, \phi=150^\circ$), corresponding to a planar rotation within the $i$-$s$ plane. The angular momenta for band B orientate at ($\theta=75^\circ, \phi=45^\circ$) and ($\theta=75^\circ, \phi=135^\circ$) due to quantum fluctuation. The profiles here are in accordance with the conclusion obtained from the $K$-distribution.

From Figs. \ref{Kdistribution} and \ref{tiltedangles}, the chiral geometry in the symmetry-restored states is illustrated by the $K$-distributions and the profile $\mathcal{P}(\theta,\phi)$ of the angular momentum.

In summary, the chiral modes are investigated with the AMP approach by diagonalizing the pairing-plus-quadrupole Hamiltonian in a symmetry restored basis. The chiral features in $^{128}$Cs \cite{Cs128-2006}, including the near degeneracy between the partner bands, the similarity of the $B(E2)$, and the staggering of the $B(M1)$, are reproduced without any free parameter. The challenge to extract the chiral geometry of the angular momentum in the AMP approach is overcome by respectively treating the components of the angular momentum in the intrinsic frame and the Euler angles as generator coordinates. The $K$-distribution in the intrinsic frame and the profile $\mathcal{P}(\theta,\phi)$ for orientation of the angular momentum are obtained, and the chiral geometry in the symmetry-restored states is illustrated. It should be noted that the AMP approach for chiral modes here can be generalized and combined with the state-of-art mean field approaches \cite{CDFT2016}. Applications for other exotic rotational modes such as the wobbling mode \cite{Bohr1975} are also possible.

\section*{Acknowledgements}

We thank Z. C. Gao for his help and discussion in the implementation of the computing codes, and X. H. Wu for his help in preparing Figure \ref{figangles}. Stimulating discussions with Z. C. Gao, F. Pan, Y. Sun, S. Y. Wang and other paticipants in the workshop ``Nuclear spontaneous symmetry breaking and its experimental signals" are acknowledged. This work was partly supported by the Chinese Major State 973 Program No. 2013CB834400, the National Natural Science Foundation of China (Grants No. 11335002, No. 11375015, No. 11461141002, and No. 11621131001), and the China Postdoctoral Science Foundation under Grants No. 2015M580007, No. 2016T90007 and No. 2017M610668.

%\bibliography{mybib}

%\end{CJK}

\end{document}